# Pulsed laser deposition of the LaVO$_4$:Eu,Ca nanoparticles on glass and silicon substrates

O. Chukova[1]*, S.A. Nedilko[1], S.G. Nedilko[1], T. Voitenko[1], M. Androulidaki[2], A. Manousaki[2], A. Papadopoulos[2], K. Savva[2], E. Stratakis[2]

1 - Taras Shevchenko National University of Kyiv, Volodymyrska Str., 64/13, Kyiv 01601, Ukraine.

2 - Institute of Electronic Structure & Laser (IESL) of Foundation for Research & Technology Hellas (FORTH), Heraklion 711 10 Crete, Greece

**Abstract**

Thin films from the LaVO$_4$:Eu,Ca nanoparticles were successfully grown by pulsed laser deposition method on glass and silicon substrates for the first time. Morphology and thickness of the films depend on a type of substrate and a number of pulses. The films are of 27 – 220 nm thickness and formed by very small particles (up to 20 nm) and also can contain single nanoparticles with dimensions 40 – 60 nm and sometimes agglomerates of nanoparticles. Spectral properties of the samples have been investigated and discussed. The vanadate films deposited on the silicon substrates lead to appearance of antireflection properties in the visible range. Luminescence spectra of the investigated films consist of narrow lines caused by f-f transitions in the Eu$^{3+}$ ions. For the samples on glass substrates the wide bands of glass emission are also contributed to the spectra. The optimal experimental conditions those allowed to obtain films promising for applications as luminescent converters are considered.

------------------

* Corresponding Author: Chukova Oksana Volodymyrivna, Ph.D.

Taras Shevchenko National University of Kyiv, Ukraine

E-mail: chukova@univ.kiev.ua

## 1. Introduction

Development of composite films with oxide nanoparticles have attracted increased research interest in relation with their applications for optical material science needs, in particular, such materials are used for downshifting of short wavelength part of spectra of incident light to the rage of spectral sensitivity of silicon solar cells and for transformation of violet and blue LED radiation into white light (Ende et al., 2009; Lewis and Nocera, 2006; Nakajima et al., 2008; Biswas et al., 2017). Currently, the most commercially-used white LEDs include luminescent converters based on the YAG:Ce particles dispersed in a polymeric (silicon) matrix (Tucureanu et al., 2015). However, polymer matrices are quickly aging under actions of light and heating (Yousif and Haddad, 2013). Attempts to solve this problem have drawn attention to creation of light transformers on the base of inorganic oxide materials (Li and Deun, 2018, Huang et al., 2013; Nedilko et al., 2015, 2016; Shinde et al., 2014).

Vanadate compounds are widely used matrices for luminescent rare earth (RE) ions dopants as they satisfy high efficiency of excitation of activator through the matrix. A wide range of possible applications of vanadate nanoparticles have attracted significant research efforts to development of new vanadate compositions with improved characteristics depending on requirements of various practical tasks (Zhu et al., 2018; Chornii et al., 2016; Biswas et al., 2017; Venkatesan et al., 2018; Li and Van Deun, 2018). Studying the $Eu^{3+}$-activated $LaVO_4$ vanadate nanoparticles, we have achieved increase of emission intensity in 10 times using various synthesis conditions (Chukova et al., 2013, 2015, 2018). The best intensity characteristics were achieved for the $La_{1-x}Eu_xVO_4$ nanoparticles synthesized by sol-gel method. Also we have developed vanadate nanoparticles with enhanced light harvesting from violet spectral range using heterovalence substitutions of RE ions with alkali earth cations. The next important task is to save high optical characteristics of oxide nanoparticles at their deposition on various substrates.

Thin films of oxide phosphors can be prepared by various methods such as chemical vapor deposition, sol-gel deposition, spin-coating, reactive sputtering, molecular beam epitaxy, liquid phase epitaxy, etc. (Higuchi et al., 2011; Inada et al., 2018; Malashkevich et al., 2016; Zorenko et al., 2005). A good way to save luminescent properties of vanadate nanoparticles is pulsed laser deposition (PLD) of films (Wicklein et al., 2012). This method doesn't require nor any matrices for dispersion of nanoarticles like a polymeric (silicon) matrix for the YAG:Ce particles used in WLEDs neither heavy metal solvents as in liquid phase epitaxy method. Also the PLD method doesn't need high temperatures and additional chemical reagents for synthesis (Savva et al., 2017). All these factors can effect on luminescent efficiency of the RE-activated nanoparticles. So, we suppose that application of the PLD method could satisfy a better preservation of enhanced optical properties of the vanadate nanoparticles in the films.

The PLD process for the $LaVO_4$ compounds was only episodically reported previously (Higuchi et al., 2011; Inada et al., 2018). The PLD films of orthovanadates were studied more intensively for $BiVO_4$ in connection with photocatalysis applications (Rettie et al., 2014; Jeong et al., 2017; Venkatesan et al., 2018) and for luminescent $YVO_4$:Eu (Kang et al., 2001; Kim et al., 2004; Zhu et al., 2014) and for $GdVO_4$:Nd (Bae et al., 2007; Bär et al., 2006; Dong et al., 2006; Yi et al., 2006) in connections with laser applications. The aim of this work is to develop and investigate model thin films with luminescent vanadate nanoparticles grown on various substrates and to study dependencies of their properties on type of substrates and growth conditions.

## 2. Experiment

The initial luminescent nanoparticles for targets preparation have $La_{1-x}Eu_xVO_4$ and $La_{1-x-y}Eu_xCa_yVO_4$ compositions. They were synthesized from calculated stoichiometric amounts of $La(NO_3)_3$, $Eu(NO_3)_3$, $NH_4VO_3$, $Ca(NO_3)_2$ precursors by sol-gel method. Nitrate solutions of the corresponded metals with a precisely defined concentration were poured into chemical utensils due to the calculated ratios. The pH was adjusted to 7.0-8.0 by ammonia solution. Then, solution of ammonium metavanadate was added. The precipitate was dissolved with a solution of citric acid in the ratio of the starting materials. The solution was concentrated by slow evaporation at 80-90 °C before formation of a gel, from which then a fine-grained powder was made and was calcined for 5 hours at 680 °C and carefully homogenized in an agate mortar. More details on sol-gel synthesis of the doped $LaVO_4$ nanoparticles can be found in (Chukova et al., 2017, 2019, a,b).

The synthesized nanoparticles were examined by XRD, SEM and spectroscopy methods. The obtained powders were grinded, homogenized and then pressed into tablets of 15 mm diameter and 2 - 3 mm thickness. Deposition was carried out in vacuum camera using KrF excimer laser with $\lambda_{gen}$ = 248 nm, 10 Hertz pulse repetition and laser fluency of 2 J/cm$^2$ on the target.

The substrates were chosen taking into account possible practical applications of the developed luminescent films. There are glass and silicon substrates. The latter are amorphous silicon plates those are used for production of silicon solar cells. The substrates (glass or silicon) were heated to 300 ºC before and at the deposition.

The microstructure of the samples was studied with a scanning electron microscope (SEM) JEOL JSM-7000F. Thin film metrology was carried out using a tabletop reflectometer (FTPadv, SENTECH Instruments GmbH) coupled to an optical microscope.

Reflectance spectroscopy of the samples was performed using Perkin Elmer Lambda 950 spectrometer. The powder samples were pressed in sample holder and then spectra of diffuse reflection were measured. In the used mode, the monochromator is placed before the samples and all light reflected from the powder samples are collected by photometric sphere. Luminescence spectra excited with 325 nm laser were registered using ACTONi (500) monochromator with grating 150 grooves/mm (blaze @ 300 nm), slit on 50 micron and liquid N$_2$ - cooled CCD camera.

### 3. Results and discussion

#### 3.1. Morphology of the films

The La$_{1-x}$Eu$_x$VO$_4$ films applied with 1000 and 2000 pulses on glass substrates have view of very light and simply light grey homogenous smoke on the glass, respectively (Fig.1). The La$_{1-x-y}$Eu$_x$Ca$_y$VO$_4$ films are characterized by additional color semitones of this grey-smoke look those are depended on point of view and lightening (Fig. 2). In despite of the homogenous smoke view of the films at usual observation conditions, the SEM images of the samples have shown that the films contain the separated particles and their small agglomerates those are spread on the glass surface (Fig. 1, 2, b). The detailed views (Fig. 1, 2, d) prove that the applied particles and small agglomerates by their sizes and forms are related with the nanoparticles used for the PLD. Comparing morphology of the films deposited on glass substrates with different number of pulses (1000 or 2000), we have not observed essential differences in character of distribution of the particles dependently on number of pulses. The last influences only density of the particles in the films. In despite of the films in their nanoscale structures contain the separated particles and their agglomerates homogenously distributed on the substrates' surfaces, at usual conditions (by eye) they are looking as complete films. The scrapers specially put on the film on glass substrate clearly demonstrates it (Fig. 2, a). Therefore, we have supposed that the films are formed by grains those are to small to be recognized in the obtained SEM images. The same situation have been reported formerly for the $YVO_4$:Eu PLD films deposited on silicon substrates (Kim et al., 2004). The authors of that investigation have obtained SEM images of the $YVO_4$:Eu PLD films those are very similar to the reported here, thus we can use their results to estimate sizes of grains in our films. Kim et al. have supposed that their films are formed by very small grains those cannot be detected by the used SEM facility. After several steps of annealing they achieved increase of grains in their films and have observed and estimated the grains. Noted, that investigated in our work films are thermally untreated and their SEM images are similar just with SEM images of thermally untreated films of (Kim et al., 2004). We have estimated grain sizes in our films using comparison of our results with the results of (Kim et al., 2004). From this comparison, we assumed that our films are formed by grains (particles) with dimensions from ~ 10 nm for the films on glass substrates to ~ 20 nm for the films on silicon substrates.

As it was shown in our previous papers, vanadate powders used here for deposition of the films consists of single nanoparticles with average dimensions of nanoparticles near 40 – 60 nm. Thus, dimensions of the film-forming grains (particles) from 10 to 20 nm are smaller than sizes of the nanoparticles in targets. We suppose the evaporated under action of laser pulses part of the target was dispersed as small-particle aerosol. It achieves the substrate and forms the films. Instead of this, some nanoparticles and their agglomerates can leave the target and achieve the substrates in their initial forms. The $La_{1-x}Eu_xVO_4$ films applied with 1000 pulses on silicon substrates are also contain the separated particles and their agglomerates those are spread on the substrates' surfaces, but some of agglomerates can be larger compared to the films on glass substrates (Fig.3).

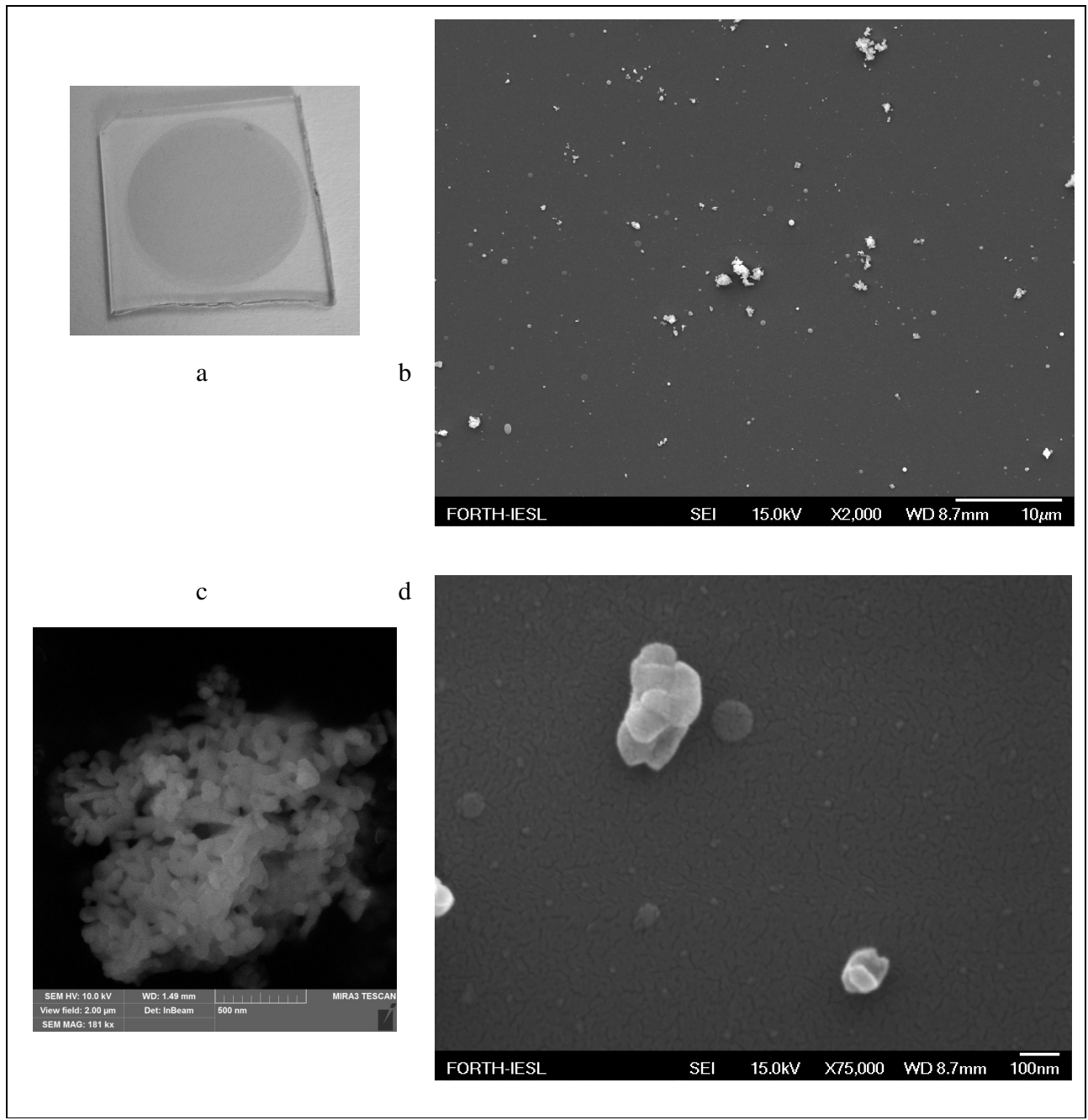

Fig. 1. Photograph (a) and SEM images (b, d) of the sample with the $La_{0.9}Eu_{0.1}VO_4$ film (1000 pulses) on glass substrate and SEM image of the nanoparticles used for the PLD (c).

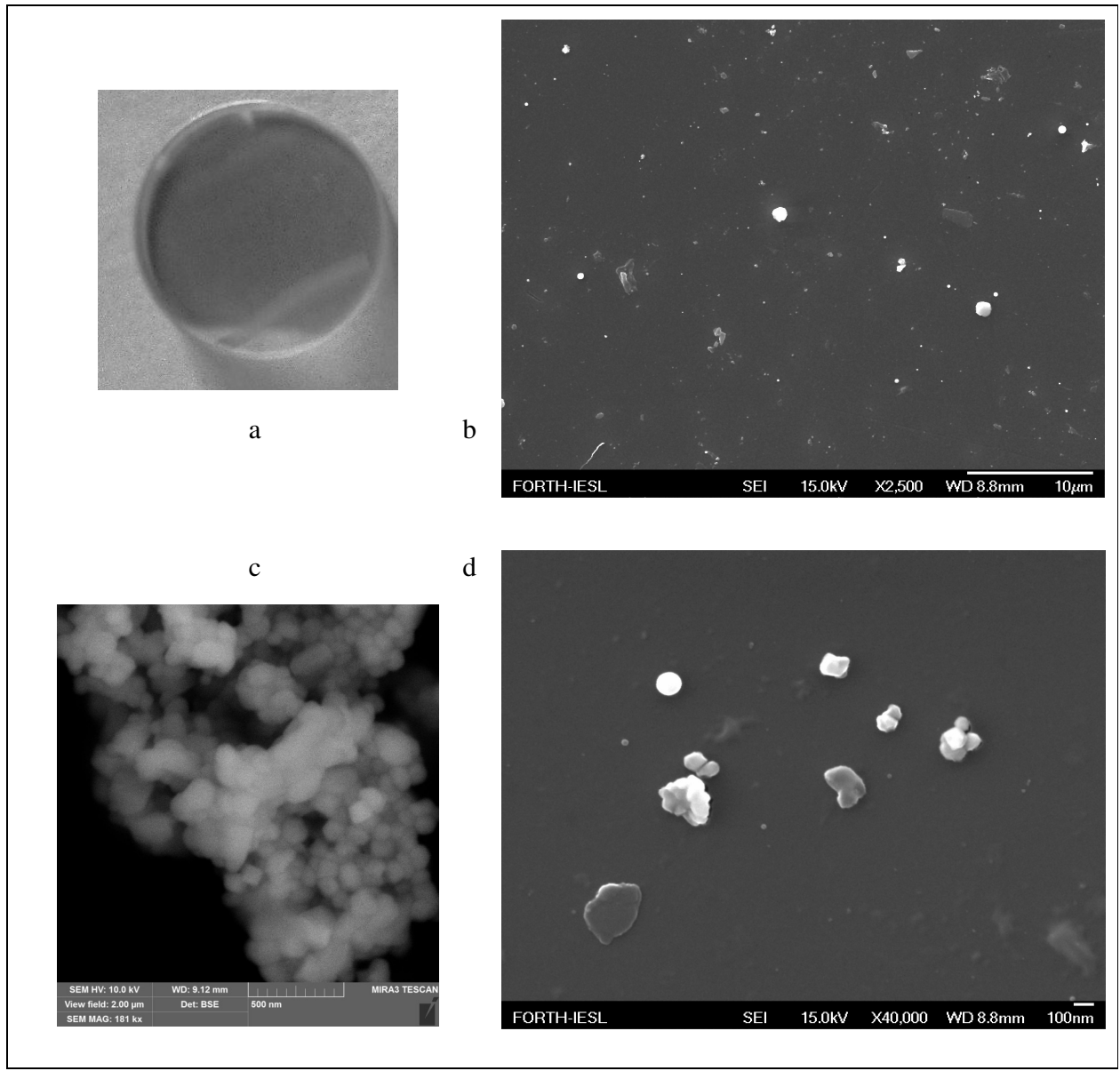

Fig. 2. Photograph (a) and SEM images (b, d) of the sample with the $La_{0.8}Eu_{0.1}Ca_{0.1}VO_4$ film (2000 pulses) on glass substrate and SEM image of the nanoparticles used for the PLD (c).

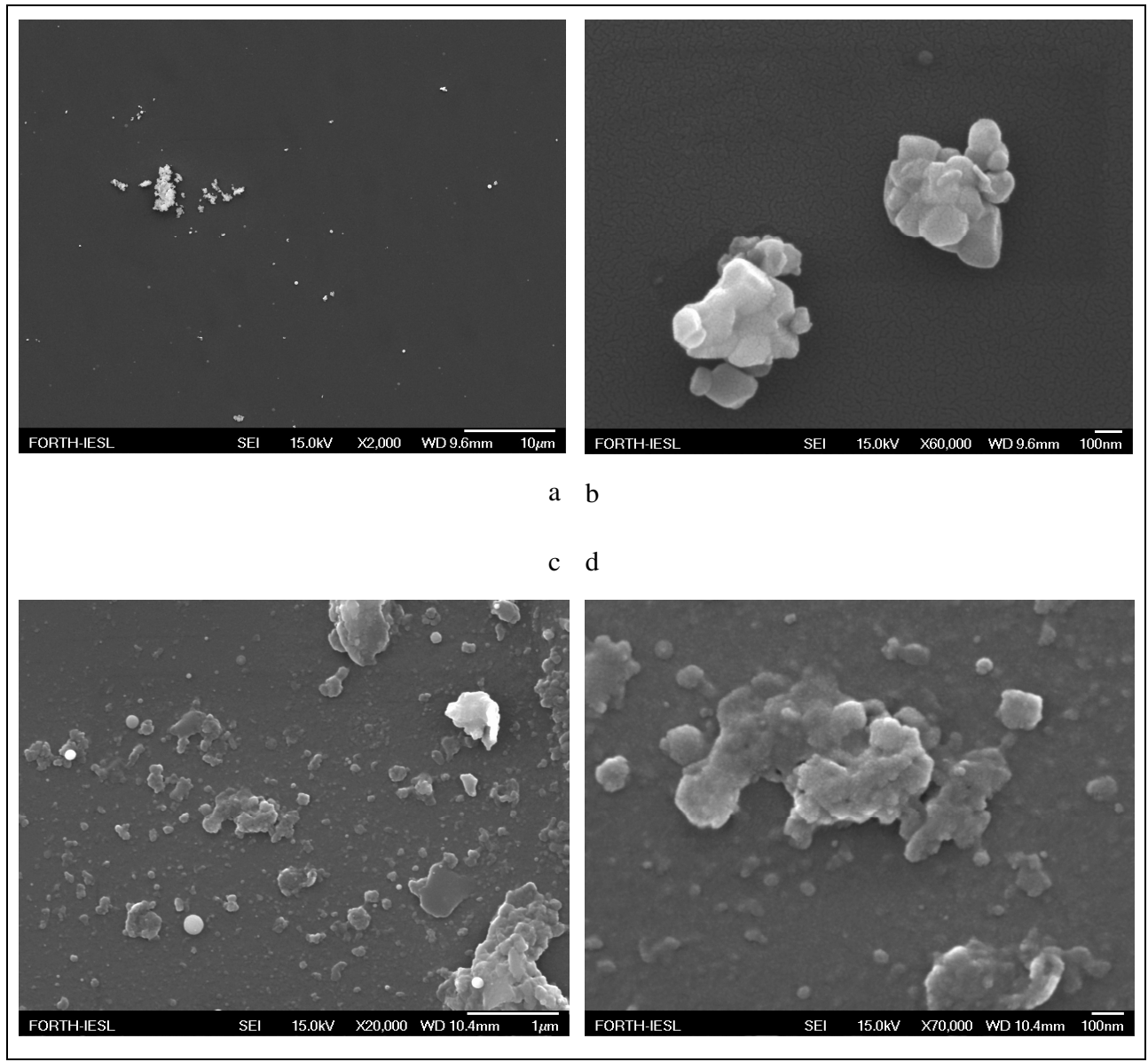

Fig. 3. SEM images the samples with the $La_{0.9}Eu_{0.1}VO_4$ film (1000 pulses) (a, b) and $La_{0.9}Eu_{0.1}VO_4$ film (2000 pulses) (c, d) on silicon substrates.

Moreover, films applied on silicon substrates are characterized by a higher density of the deposited nanoparticles. We suppose that different densities of nanoparticles in the obtained films are appeared as a result of different interaction of the nanoparticles with the used substrates. The amorphous silicon substrate is characterized by high surface roughness that satisfies better adhesion of the deposited nanoparticles. This property should be taken into account for optimization of parameters of the PLD processes. At the same time, we suppose that high degree of agglomeration of the initial nanoparticles (Chukova et al., 2017, 2019a) is responsible for appearance of separate particles and some agglomerates on the films. So, some treatment of the powders before pressing into tablets is necessary in order to obtain more homogenous films.

Thicknesses of the grown films were estimated using optical thin film metrology method based on the interference patterns in the reflectivity spectrum of the layered sample. Interference effects occurred between the radiation reflected at the boundary and at the surface of the film were investigated with optical microscope and tabletop reflectometer. The light reflected from the sample is collected by pick-up optics connected to the camera adapter of a microscope. An optical fiber guides the light to the spectrometer and the spectrum is detected by a high performance photo diode array. The measured spectrum from a layered sample shows typical interference effects those depend on the thickness and the optical properties of the materials. Therefore the film thickness can be determined using optical constants (n and k) of film material and substrate (Mochi et al., 2020). The calculations were performed by own non-commercial programs.

This investigation has shown that the applied method estimates the deposited samples as films with the certain thicknesses of nanoscale dimensions (Table 1). Thus, observed distributions of nanoparticles and their agglomerates doesn't influences on metrology characteristics of the films. The measured thicknesses were estimated as from 27 to 70 nm for the films on silicon substrates and from 180 to 220 nm for the films on glass substrates. Comparing results of SEM study and optical metrology, we should conclude that in despite of higher thickness of films on glass substrates, they contain a lower density of agglomerated inclusions. Therefore, the PLD technology allows to obtain very thin films of good quality from the vanadate nanoparticles, especially if compare with such methods as sol-gel deposition, spin-coating, reactive sputtering, and liquid phase epitaxy (Chukova et al., 2019b; Zhu et al., 2018; Malashkevich et al., 2016; Nakajima et al., 2008; Zorenko et al., 2005).

Table 1. Compositions of the films, number of pulses, types of substrates and thicknesses of the films estimated using the thin film metrology technique

|   | Film composition | Number of pulses | Substrate | Thickness, nm |
|---|---|---|---|---|
| 1 | $La_{0.9}Eu_{0.1}VO_4$ | 1000 | glass | 180 |
| 2 | $La_{0.9}Eu_{0.1}VO_4$ | 2000 | glass | 219 |
| 3 | $La_{0.8}Eu_{0.1}Ca_{0.1}VO_4$ | 1000 | glass | 206 |
| 4 | $La_{0.9}Eu_{0.1}VO_4$ | 1000 | silicon | 27 |
| 5 | $La_{0.9}Eu_{0.1}VO_4$ | 2000 | silicon | 70 |
| 6 | $La_{0.8}Eu_{0.1}Ca_{0.1}VO_4$ | 1000 | silicon | 38 |

### 3.2. Optical properties

Investigation of reflection and transmittance (for the samples on glass substrates) spectra was carried out to estimate contributions of the film layers in spectral properties of the obtained samples. The films applied on the glass substrates increase reflection of the samples in 2-3 times from 8-10 to 20-30 % in the spectral interval from 400 nm to near IR range. Increase number of pulses increases the noted effect (Fig. 4, curves, 2, 3). No additional spectral features were observed for this interval. Distinctive differences for the film samples are observed in the 250 – 400 nm spectral diapason. There are sharp edges of reflection those are centered at 300 and 325 nm for the $La_{0.9}Eu_{0.1}VO_4$ and $La_{0.8}Eu_{0.1}Ca_{0.1}VO_4$ films, respectively. The same band is observed in the transmittance spectra. These bands should be ascribed to well-known absorption transitions in the vanadate groups The same properties were reported previously for the $La_{1-x}Eu_xVO_4$ and $La_{1-x-y}Eu_xCa_yVO_4$ nanoparticles (Chukova et al. 2019a). The observed increase of reflection from the samples in the spectral interval from 400 nm to near IR range should be also assigned to contribution of the vanadate films. Diffuse reflectance spectra of the dense layers of the initial vanadate nanoparticles are characterized by reflection higher than 90 % in the noted spectral range. Increase of reflection of the glass substrate onto 10 – 20 % (from 8-10 to 20-30

%) with the films deposition means that about 10 – 20 % of incident light is reflected from the sample by the deposited vanadate nanoparticles.

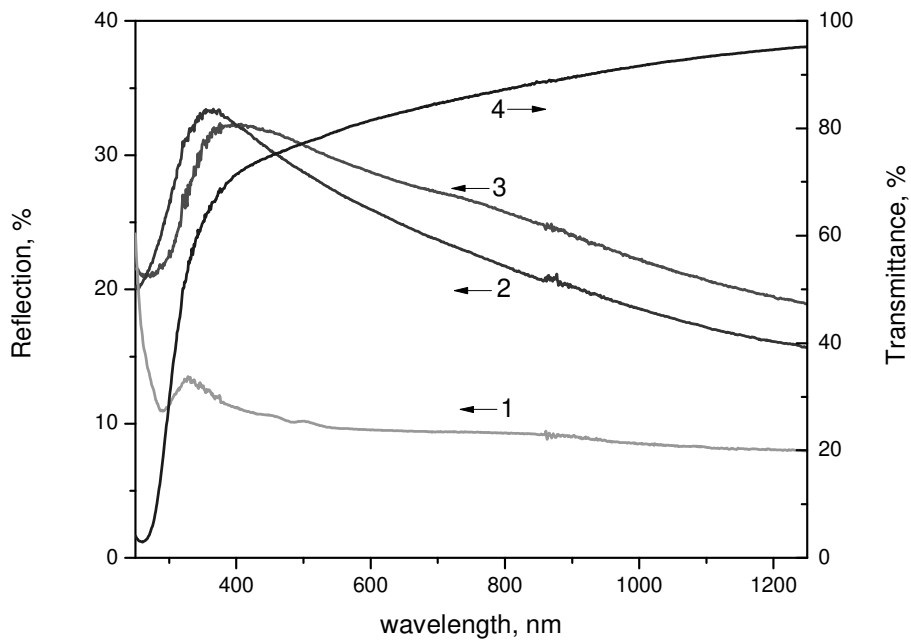

Fig. 4. Reflection (1-3) and transmittance (4) spectra of the samples with the $La_{0.9}Eu_{0.1}VO_4$ (1000 pulses) (2, 4) and $La_{0.8}Eu_{0.1}Ca_{0.1}VO_4$ (2000 pulses) (3) films on glass substrate (1)

Reflection spectra of the films on silicon substrates have sharp edge centered at 350 nm for the $La_{1-x}Eu_xVO_4$ films (Fig. 5, curves 2, 3). Difference of 50 nm in the position of sharp edge of the reflection spectra of the films obtained from same nanoparticles appears due to contribution of glass substrate in the total reflection spectra of the samples (see Fig. 4, curve 1), whereas silicon substrate doesn't essentially contribute in reflection spectra up to 300 nm (Fig. 5, curve 1). We have also registered strong decrease of total reflection of the samples with the films on silicon substrates from 90 % to 30 % observed over all the visible range. As this decrease have no any spectral features and the same for various samples, we assume that it is caused by multiple scattering followed with absorption of the incident light in the deposited films. This agrees with the above described morphology of the films those are formed by separated nanoparticles and their agglomerates.

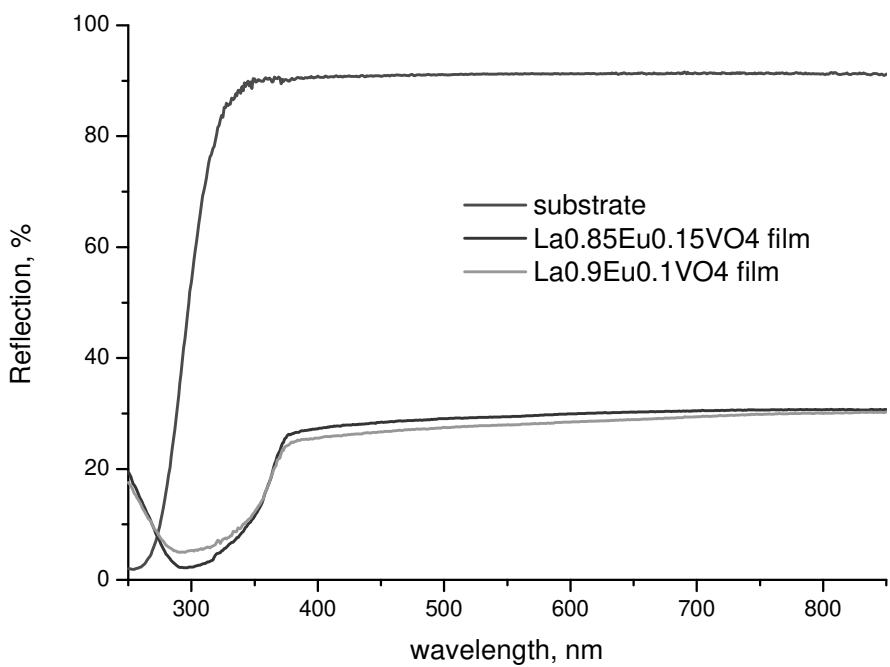

Fig. 5. Reflection spectra of the samples with the $La_{0.9}Eu_{0.1}VO_4$ and $La_{0.85}Eu_{0.15}VO_4$ films (1000 pulses) on silicon substrates.

Possibility of formation of the nanostructures with antireflection properties under action of ultrafast laser pulses was reported recently (Papadopoulos et al., 2019; Goodarzi et al., 2020). Antireflection properties of random nanostructures are well known for various natural objects (Siddique et al., 2015). We don't exclude that deposited vanadate films on silicon substrates are able to show some antireflection properties as a result of randomly nanostructured profile. Usually, production of silicon solar cells includes special antireflection coating of silicon plates with silicon nitride or titanium dioxide. Thus, the films of vanadate nanoparticles can also play role of antireflection coating on the silicon substrates.

3.3 Luminescence

Luminescence spectra of the samples on silicon substrates consist of groups of narrow lines in the 575 – 725 nm spectral range. Luminescence spectra of the samples on glass substrates consist of three wide bands at 420, 520 and 700 nm and the groups of narrow lines in the 575 – 725 spectral range are observed on the background of the wideband emission. The narrow lines in the 575 – 725 nm spectral range by their positions and intensity distributions should be assigned to electron transitions in the $Eu^{3+}$ ions. The wide bands at 420, 520 and 700 nm those are observed for the samples on glass substrates only can be caused by emission of glass substrate (Fig. 6).

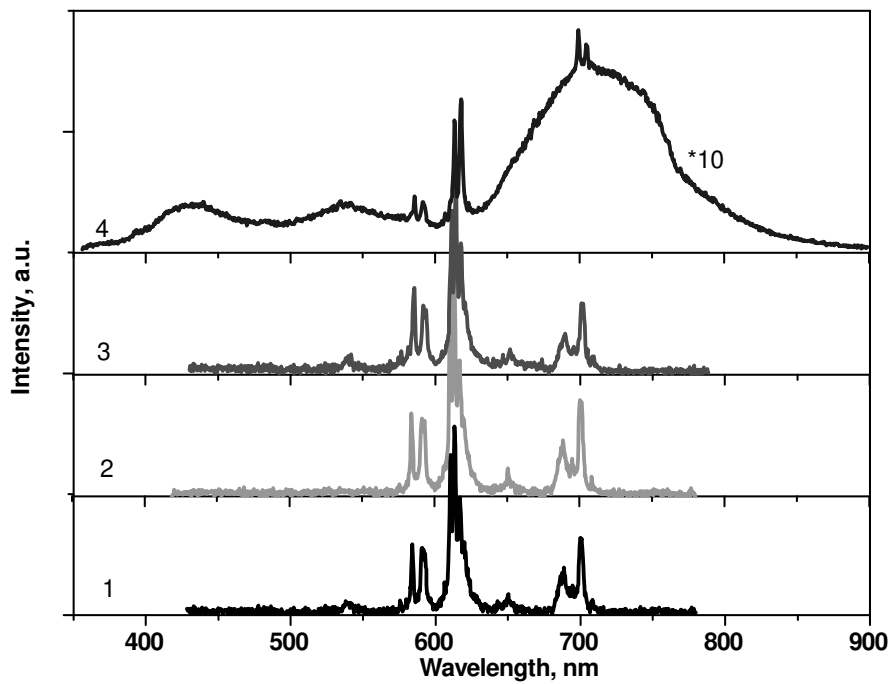

Fig.6. Luminescence emission spectra of the $La_{0.9}Eu_{0.1}VO_4$ (1000 pulses) (1) (2000 pulses) (2, 4) and $La_{0.8}Eu_{0.1}Ca_{0.1}VO_4$ (3) films on silicon (1-3) and glass (4) substrates

Comparing luminescent properties of the films on glass and silicon substrates, we should assume that intensity of emission of the $Eu^{3+}$ ions strongly depends on type of substrate. The highest intensity is observed for the $La_{0.8}Eu_{0.1}Ca_{0.1}VO_4$ films on silicon substrates. The films on glass substrates are characterized by low intensity of the $Eu^{3+}$ emission. We consider that several factors can be a cause of the observed dependency of intensity of luminescence emission on type of substrate. First of all, the films applied on silicon substrate are characterized by a higher density of the nanoparticles and their agglomerates. Secondly, the synthesized films are very thin, and part of excitation light passes through the films. Then, in a case of glass substrates this part of excitation light passes also through glass substrate and lefts the sample. As for the used amorphous silicon substrates, they have rough surfaces those lead to multiple scattering. The multiple scattering is followed with increased absorption of excitation light in the film layer that results in such a way in enhancement of efficiency of luminescence excitations.

The studied luminescent properties of the obtained films can help to find optimal synthesis conditions for possible practical applications of the films. For example, if we consider films on glass substrates for luminescent converting of violet or blue emission of LEDs, we should say that the obtained films are too thin to obtain of sufficient luminescent characteristics of films. The used experimental conditions of 1000 - 2000 pulses are not enough for this application. From the other hand, the films on silicon substrates have demonstrated very good characteristics to be used as luminescent downshifters of the incident light by solar cells. They are characterized by high intensity of emission and high efficiency of transformation of light from near UV range to red and at the same time their high transparency in the visible range will not prevent penetration of solar light to working solar cells surfaces. Therefore, it is possible to preserve enhanced optical characteristics of vanadate nanoparticles under their deposition on various substrates. But, effective excitation of very thin films is possible only for samples on silicon substrates. Taking into account that used in this investigation substrates have not obtained special treatment those are usually used for obtaining of solar cells from silicon plates (metallization, electrodes application, antireflection coating, etc.), the additional investigation is needed to estimate possibility of the proposed films to improve spectral efficiency of the end devices.

## 4. Conclusions

Thin films from the LaVO$_4$:Eu,Ca nanoparticles were successfully grown by pulsed laser deposition method on various glass and silicon substrates for the first time.

Morphology and thickness of the films depend on a type of substrate and a number of pulses. The obtained films are characterized by thicknesses from 27 to 220 nm. The films are formed by very small particles (up to 20 nm) and also can contain single nanoparticles with dimensions 40 – 60 nm and sometimes agglomerate of nanoparticles.

Spectroscopy properties of the obtained samples have been investigated. The deposited vanadate films on silicon substrates have demonstrated antireflection properties as a result of laser-induced random nanostructured profile. Luminescence spectra of the investigated films consist of narrow lines caused by f-f transitions in the $Eu^{3+}$ ions. Intensity of the $Eu^{3+}$ emission is considerably higher for the films deposited on silicon substrates. For the samples on glass substrates the wide bands of glass emission are also contributed in the spectra. The used experimental conditions are not enough to obtain films on glass substrates with luminescent characteristics of the films sufficient for their applications, whereas the films deposited on silicon substrates have demonstrated promising antireflection and luminescent characteristics.

## 5. Acknowledgments

This work has received funding from Ministry of Education and Science of Ukraine (University's Science Grant Program and Bilateral Grant Program) and from the EU-H2020 research and innovation program under grant agreement No 654360 having benefited from the access provided by Institute of Electronic Structure & Laser (IESL) of Foundation for Research & Technology Hellas (FORTH) in Heraklion, Crete, Greece and Paul Scherrer Institut (PSI), Villigen, Switzerland within the framework of the NFFA-Europe Transnational Access Activity. Authors thanks Dr. Dimitrios Kazazis (PSI) for his kind support with the thin film metrology experiments (thicknesses of the films).

**Conflict of interest statement.**

On behalf of all authors, the corresponding author states that there is no conflict of interest.